\documentclass[aps,prl,twocolumn,a4paper,10pt,notitlepage,footinbib,superscriptaddress,showpacs]{revtex4-1}%
\usepackage[english]{babel}
\usepackage[T1]{fontenc}
\usepackage{endnotes}
\usepackage{amssymb,amsmath,amsfonts}
\usepackage{textcomp}
\usepackage{graphicx,color}
\usepackage[dvipsnames]{xcolor}
\usepackage[utf8x]{inputenc}
\usepackage{float}

\usepackage{tabularx}

\renewcommand{\vec}[1]{\boldsymbol{\mathrm{#1}}}%

\newcommand{\derpar}[2]{\frac{\partial#1}{\partial#2}}
\newcommand{\derfunc}[2]{\frac{\delta#1}{\delta#2}}

\newcommand{\restrdelta}{\left(\left.\Delta\right|_\gamma\right)}

\newcommand{\ateq}[1]{{#1}^\star}
\newcommand{\set}[1]{\left\{#1\right\}}
\newcommand{\norm}[1]{\left\|#1\right\|}
\newcommand{\braket}[1]{\left\langle#1\right\rangle}

\begin{document}

\title{Magnetic polymer melts and epigenetic phase separation} 
\title{Magnetic Polymer Models for Epigenomic Organisation and Phase Separation} 

\vspace{-0.2 cm}
\author{Davide Col\`i$^*$ }
\affiliation{Dipartimento di Fisica e Astronomia and Sezione INFN, Università degli Studi di Padova, I-35131 Padova, Italy}
\author{Davide Michieletto$^*$}
\affiliation{SUPA, School of Physics and Astronomy, University of Edinburgh, Edinburgh EH9 3FD, United Kingdom}
\author{Davide Marenduzzo}
\affiliation{SUPA, School of Physics and Astronomy, University of Edinburgh, Edinburgh EH9 3FD, United Kingdom}
\author{Enzo Orlandini}
\affiliation{Dipartimento di Fisica e Astronomia and Sezione INFN, Università degli Studi di Padova, I-35131 Padova, Italy}

\date{\today}

\begin{abstract}
\vspace{-0.3 cm}
\textbf{The genetic instructions stored in the genome require an additional layer of information to robustly determine cell fate. This additional regulation is provided by the interplay between chromosome-patterning biochemical (``epigenetic'') marks and three-dimensional genome folding. Yet, the physical principles underlying the dynamical coupling between three-dimensional genomic organisation and one-dimensional epigenetic patterns remain elusive. To shed light on this issue, here we study by mean field theory and Brownian dynamics simulations a magnetic polymer model for chromosomes, where each monomer carries a dynamic epigenetic mark. At the single chromosome level, we show that a first order transition describes the unlimited spreading of epigenetic marks, a phenomenon that is often observed {\it in vivo}. At the level of the whole nucleus, experiments suggest chromosomes form micro-phase separated compartments with distinct epigenetic marks. We here discover that for a melt of magnetic polymers such a morphology is thermodynamically unstable, but can be stabilised by a non-equilibrium and ATP-mediated epigenetic switch between different monomer states.
}
\end{abstract}

\maketitle

\vspace{-1.3 cm}
\paragraph{Introduction}
Each cell in our body contains the same DNA and hence it carries the same genetic information; yet, cells in different tissues possess distinct identities that are robustly inherited following multiple rounds of cell division~\cite{Alberts2014,Cortini2016}. 
Thus, cellular fate cannot be directed by genetic cues alone and it requires an additional layer of information involving 3D genome organisation~\cite{Cavalli2013,Ciabrelli2017,Boettiger2016} and tissue-specific ``epigenetic'' patterns~\cite{Moazed2011,Angel2011,Laprell2017,Dodd2007,Berry2017, Berry2015}. The latter consist of biochemical tags that are deposited along the genome and on histones -- the proteins in charge of packaging DNA into chromatin~\cite{Alberts2014,Cortini2016}. The interplay between spatial genome organisation and epigenetic patterns guides the tissue-specific selection of which genes will be translated into proteins, in turn determining cellular identity~\cite{Gilbert2004cell,Boettiger2016,Pombo2015, Cavalli2013,Probst2009}. One of the outstanding problems in biophysics is to understand how genome organisation and epigenetic patterns are linked to each other dynamically and what are the physical principles through which they regulate genome functionality and cellular memory~\cite{Michieletto2016prx,Michieletto2018nar,Haddad2017,Jost2018, Pancaldi2016,Erdel2013,Teif2015}. 

To shed light on this issue here we introduce and study, both analytically and numerically, models of the genome where its 3D spatial organisation is coupled to a dynamically evolving epigenetic field~\cite{note1}. These models describe each chromosome as a magnetic polymer whose monomers encode (epigenetic) states which can change over time; they are therefore in the same universality class of annealed copolymers without global conservation laws~\cite{Garel1988}. This model is markedly different from previous works on annealed copolymers with conserved number of elements in each state~\cite{Grosberg1984,Dormidontova1992,Sfatos1997}, and can be seen as a generalisation of the 1D Ising (or Potts) system where the substrate is allowed to diffuse in 3D space~\cite{Michieletto2016prx,Michieletto2018nar, Michieletto2017scirep}. We combine analytical mean-field theories with Brownian Dynamics (BD) simulations to simultaneously map the distribution of epigenetic marks and the 3D genomic arrangement within the cell nucleus. Together they describe the nuclear ``epigenomic'' organisation that can be directly compared with experiments~\cite{Cavalli2013,Cremer2015}. 

At the single chromosome level, our magnetic polymer undergoes a first order transition between a swollen, epigenetically disordered fibre and a compact, epigenetically ordered one. Dynamically, the model generically predicts uncontrolled growth of the dominating epigenetic mark, reminiscent of the process through which transcriptionally repressed chromatin is often seen to spread {\it in vivo}, e.g., in X-chromosome inactivation~\cite{Pinter2012,Nozawa2013} or position-effect-variegation~\cite{Moazed2011,Hathaway2012}. At the whole nucleus level, a melt of magnetic polymers can initially phase separate into multiple thermodynamically metastable epigenetic domains; these though evolve into a single domain at large times. Introducing a local non-equilibrium epigenetic switch between an epigenetically active and an inactive state -- mimicking ATP-dependent chromatin remodelling processes which modify chromatin accessibility to the deposition of biochemical marks -- arrests the phase separation and ordering kinetics and yields micro-phase separation of the genome into multiple epigenetic domains, reminiscent of those observed in the cell nucleus~\cite{Rao2014,Cremer2015,Cavalli2013,Strom2017,Larson2017}. 
\paragraph{Single Chromosome}
To describe the equilibrium properties of a single chromosome fibre with a fluctuating epigenetic profile 
we consider an $N$-step self-avoiding walk (SAW) on a lattice with coordination number $z$ where each vertex displays an epigenetic state $q$. The partition function of the model reads
\begin{equation}
{\cal{Z}} = \sum_{SAW} \sum_{q} \exp \left [ -\frac{\beta}{2} \sum_{i,j=1}^N \Delta_{r_i,r_j}J(q_i,q_j)\right ]
\label{part_funct}
\end{equation}
where $1/\beta=k_B T$ and $\Delta_{r,r'} = 1$ if $(r,r')$ are nearest-neighbour on the lattice (and 0 otherwise) thus restricting the interaction to 3D proximal segments.

For simplicity, we limit our model to three possible epigenetic states~\cite{Dodd2007} ($q_i\in\{-1,0,1\}$) and define $J(q_i,q_j) = -\epsilon$ if $q_i = q_j = \pm 1$  and 0 otherwise. With this choice we implicitly assume that two marks are self-attractive ($q_i= \pm 1$) while the third ($q_i=0$) is neutral (or unmarked)~\cite{Dodd2007,Michieletto2016prx}. This choice is also motivated by biological consideration, as we can assume that the non-neutral polymer states are associated with read-write protein complexes that can bridge polymer segments bearing the same epigenetic mark while ``infecting'' spatially neighbouring segments with the same mark~\cite{Dodd2007,Berry2017}. Both processes are captured by the same energy term and are akin to ferromagnetic interactions that align 3D proximal Ising or Potts spins, or bring them together when already aligned~\cite{Garel1988}.


Eq.~\eqref{part_funct} can be solved within a mean field approximation~\cite{Garel1988,Garel_et_al_EPJB_1999a} for an Ising-like model on a SAW (see SI for details). This approximation leads to the free energy density
\begin{eqnarray} 
\dfrac{f}{T} &=& -\log \left (\frac {z}{e} \right ) + \frac{1-\rho}{\rho}\log (1-\rho) - \frac{\alpha}{10} \rho\nonumber \\
    &+& \frac{9}{10\alpha} \frac{\phi^2}{\rho} - \ln \left (e^{6\phi/5}+2e^{-3\phi/10} \right )  \, ,
\label{free_energy_single}
\end{eqnarray}
where $\alpha\equiv \beta \epsilon z$ is the interaction parameter strength, $\rho \equiv N/V$ the chromosome density (as the chain is confined into a box of volume $V$) and $\phi$ is an epigenetic field (here modelling global epigenetic ordering). In analogy to ferromagnetic systems~\cite{Garel1988}, we can identify $\phi$ as the average magnetisation of the system (see SI). 

\begin{figure}[!t]
\centering
\includegraphics[width=0.4\textwidth]{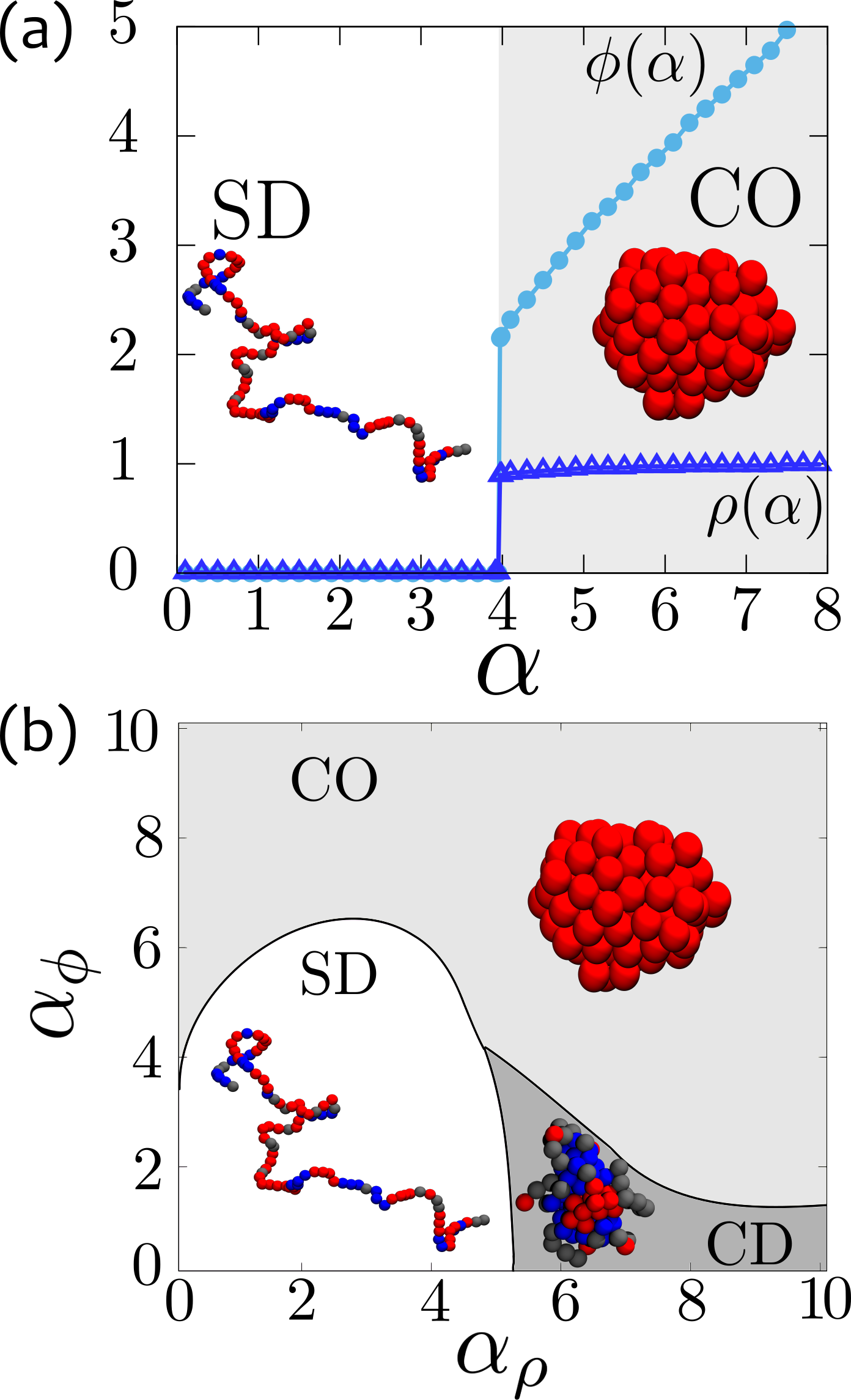}
\vspace{-0.3 cm}
\caption{\textbf{Phase Diagrams at the single chromosome scale.} \textbf{(a)} Equilibrium phase diagram of a magnetic ``epigenetic'' polymer described by the free energy in Eq.~\eqref{free_energy_single}. The system undergoes a first order transition (marked by a discontinuity in the order parameters) at $\alpha = \epsilon/k_B T \simeq 3.96$ between a swollen-disordered (SD) and a compact-ordered (CO) phase. \textbf{(b)} Non-equilibrium phase diagram obtained integrating Eqs.~\eqref{crit_nondyn_eq_single} in the parameter space ($\alpha_\rho$, $\alpha_\phi$). Insets: snapshots of representative configurations from BD simulations (see SI for details). }
\vspace{-0.5 cm}
\label{fig:singlechain}
\end{figure}

By minimizing Eq.~\eqref{free_energy_single} with respect to $\rho$ and $\phi$ one obtains the equilibrium phase diagram (see Fig.~\ref{fig:singlechain}) where we distinguish two phases. At low $\alpha$ the system is in a \emph{swollen-disordered} phase (SD): the chain is extended ($\rho=0$) and heterogeneously coloured ($\phi^2=0$). At large $\alpha$ we find a \emph{compact-ordered} phase (CO) where the chain is crumpled ($\rho\ne 0$) and nearly uniformly coloured $(\phi^2 > 0)$. The discontinuous jumps of the order parameters $\rho$ and $\phi$ at the transition point ($\alpha_c\simeq 3.96$) signal a first order transition between these two regimes~\cite{Michieletto2016prx,Michieletto2017nar}. In Fig.~\ref{fig:singlechain} we also report snapshots of representative configurations from BD simulations of a corresponding polymer model where the Langevin dynamics of the polymer backbone is coupled to a Monte-Carlo annealing procedure that evolves the states of the polymer beads (see~\cite{Michieletto2016prx} and SI for details). The first order nature of the transition, as noted in~\cite{Michieletto2016prx}, provides a mechanism to endow memory to a global epigenetic state.

A relaxation dynamics for $\rho$ and $\phi$ can be written down starting from Eq.~\eqref{free_energy_single} in terms of two coupled ``Model A'' equations~\cite{ChaikinLubensky} as both fields are not conserved at the single chromosome level. [Here $\rho$ should be understood as the density of beads within the smallest box containing the polymer chain]. Such equations read
\begin{eqnarray}\label{crit_nondyn_eq_single}
\frac{\partial \phi ({\bf r},t)}{\partial t} &=& -\Gamma_{\phi} \left( \frac{9\phi}{5\alpha_{\rho}\rho}+\frac{3}{e^{3\phi/2}+2} -\frac{6}{5}\right ) + \kappa_{\phi}\nabla^2 \phi,  \\ 
\frac{\partial \rho ({\bf r},t)}{\partial t} &=& \Gamma_{\rho} \left(\frac{9\phi^2}{10\alpha_{\rho}\rho^2}+ \frac{\alpha_{\rho}}{10}+\frac{\ln(1-\rho)}{\rho^2}+\frac{1}{\rho}\right ) + \kappa_{\rho}\nabla^2\rho \, , \notag
\end{eqnarray}
where $\Gamma_{\rho/\phi}$ and $\kappa_{\rho/\phi}$ are mobilities and surface tension-like coefficients, respectively.  In Eqs.~\eqref{crit_nondyn_eq_single} we decouple $\alpha$ into two independent parameters affecting the dynamics of the polymer ($\alpha_\rho$) and of the epigenetic field ($\alpha_\phi$) separately. Note that the case $\alpha_\rho \neq \alpha_\phi$ leads to non-equilibrium dynamics as these equations no longer derive from a free energy. By numerically integrating Eqs.~(\ref{crit_nondyn_eq_single}) we obtain the non-equilibrium phase diagram shown in Figure~\ref{fig:singlechain}(b). We discover a new phase that is absent in equilibrium ($\alpha_\rho=\alpha_\phi$), featuring a crumpled and epigenetically disordered (CD) polymer. Yet, within the mean field approximation, we do not observe the swollen-ordered (SO) phase seen in BD~\cite{Michieletto2016prx,Michieletto2017nar}. 

A single magnetic ``epigenetic'' polymer therefore exists in one of three phases in steady state, each reminiscent of a biologically relevant configuration. The SD phase models the conformation of a chromosome exiting mitosis, when epigenetic patterns and 3D folding are not yet established~\cite{Cavalli2013,Nagano2017}. The CO phase resembles the ``Barr body'' into which the inactive X-chromosome folds in female mammalian cells~\cite{Pinter2012}. This is a dense globular structure which is homogeneously marked with a repressive epigenetic state~\cite{Nozawa2013}. Finally, the CD phase is akin to inert chromatin which experimental contact maps suggest is compact~\cite{Sexton2012,Rao2014}, yet has no clear epigenetic signature~\cite{Sexton2012,Filion2010,Saksouk2015}. 

Our theory also offers a framework within which to understand the spreading of repressive marks (heterochromatin) in X-chromosome inactivation or in other position-effect-variegation where a transcriptionally silent domains spreads onto a nearby gene, switching off its expression~\cite{Henikoff2008,Hathaway2012}. In our model the spreading occurs via a $t^{1/2}$ growth~\cite{ChaikinLubensky} when both epigenetic states are equally likely, whereas if one is favoured we expect linear Fisher-like growth~\cite{Murray:2002}.


\paragraph{Whole nucleus}
At the scale of the entire nucleus (volume $V$) we assume that chromosomes are initially homogeneously filling the space. The overall density of chromatin $n_0 \equiv N/V$ (where $N$ is now the total length of the genome) is conserved, and the system can be described as a melt of magnetic polymers.

\begin{figure}[t]
\centering
\includegraphics[width=0.45\textwidth]{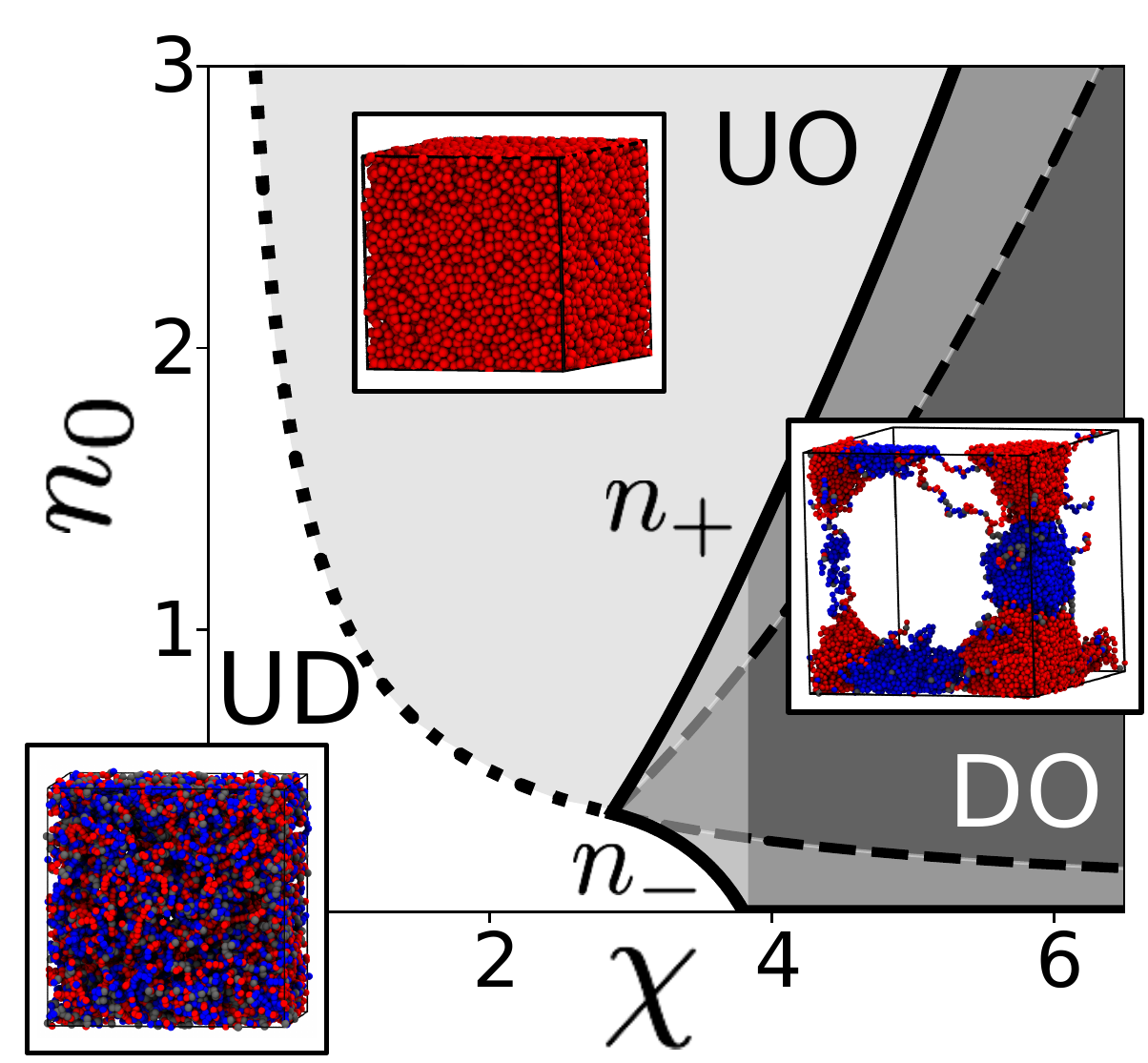}
\caption{\textbf{Phase Diagram at the Nuclear Scale} Equilibrium phase diagram of a melt of magnetic polymers obtained by using the common tangent construction on the free energy Eq.~\eqref{eq:FMelt}. The three equilibrium phases are: (UD) uniform ($n=n_0$) and epigenetically disordered ($m^2=0$); (UO) uniform ($n=n_0$) and epigenetically ordered ($m^2>0$); (DO) demixed and epigenetically ordered ($n=n_+$, $m^2>0$ and  $n=n_-$,$m^2=0$). A fourth partially-demixed ordered (PDO) phase is characterised by weaker variations in density $(n_->0)$ and denoted by a white shading within the DO phase. The dotted line marks the critical value of the coupling $\chi_c(n_0)$, the solid lines identify the boundaries of the coexistence region (binodals) and the dashed lines identify the spinodal region where the uniform solution is linearly unstable~\cite{Fosado2017,ChaikinLubensky}. Insets report representative snapshots from Brownian Dynamics simulations of dense solution of magnetic polymers (see SI for details).} 
\vspace{-0.6 cm}
\label{fig:pd_melt}
\end{figure}

A minimal free-energy density describing the equilibrium properties of this model is  
\begin{equation}
\beta f =  a m^2 + b m^4 + c n^2 + d n^3 -\chi m^2 n \,
\label{eq:FMelt}
\end{equation}
where the fields $n(\vec{x}, t)$ and $m(\vec{x}, t)$  are the local chromosome density distribution and the average epigenetic marks respectively. The terms in Eq.~\eqref{eq:FMelt} can be justified as follows: (i) the magnetisation field should not explicitly break its intrinsic $\mathcal{Z}_2$ symmetry (if both marks are equally likely); (ii) the density field should be described by a standard virial expansion for non-ideal gases; (iii) the minimal coupling $\chi m^2 n$ should capture the interplay between chromatin folding ($n>0$) and epigenetic ordering ($m^2 > 0$). For convenience and without lack of generality we set $a>0,b>0,c>0$, $d>0$ and $\chi = \chi(T) >0$. The equilibrium phase diagram (see Fig.~\ref{fig:pd_melt}) is obtained by first minimising Eq.~\eqref{eq:FMelt} with respect to the non conserved field $m$, i.e. $\left. \partial f / \partial m \right|_{m^*}= 0$, and then by analysing the resulting $f(m^*,n)$ as a function of the conserved field $n$, via a common tangent construction~\cite{Fosado2017,Matsuyama2002}.

For small values of $\chi$ the system is in a uniform ($n=n_0$) and epigenetic disordered phase ($m=0$) (UD) (no epigenomic domains). Upon increasing the overall density $n_0$ (keeping $\chi\le \chi_c (n_0)$ fixed) we find a second order phase transition to a uniform state with ordered epigenetic field $(m({\bf x})^2>0)$ (UO) (see SI). The dynamics of the UD-UO transition is characterised by long-lived bicontinous spanning domains with alternated epigenetic marks (see Suppl. Movies), similar to growing magnetic domains in Ising systems~\cite{ChaikinLubensky,Kockelkoren2002}. At large times these domains coalesce into a single system-spanning epigenetic domain (see Figs. S1,S2 and inset in Fig.~\ref{fig:pd_melt} from the BD simulations). Finally, for $\chi > \chi_c (n_0)$, we observe that the uniform state is unstable and the system phase separates into high ($n_+$) and low ($n_-$) density regions forming a demixed-ordered (DO) phase. The high-density regions are associated with strong epigenetic domains ($m^2>0$) whereas the low-density regions with neutral epigenetic signature ($m^2=0$) (see Fig.~\ref{fig:pd_melt}). This phase is contained within the binodal curves which are determined using a common tangent construction~\cite{Fosado2017,ChaikinLubensky,Matsuyama2002} (see SI). 
We also mention that close to the critical point, where the binodal lines meet, the DO phase displays weaker variations of density throughout the system, i.e. the low density phase is strictly non-zero $(0< n_- < n_+)$. We call this regime partially demixed ordered (PDO) phase (see Fig.~\ref{fig:pd_melt}). Pleasingly, the equilibrium phases obtained from the mean-field  free-energy~\eqref{eq:FMelt} are confirmed by BD simulations of a more realistic model in which the genome is described as a dense solution of magnetic polymers (see insets of Fig.~\ref{fig:pd_melt} and SI). 


Some of the observed phases are reminiscent of the epigenomic organisation seen in experiments. The UD phase (as the SD phase for a single polymer) may represent a genomic configuration upon exit from mitosis, when spatial structure and epigenetic patterns are yet to be established (although our model does not account for mitotic chromosome structure). The (P)DO phase may be associated to strongly phase-separated nuclei, for instance in retinal~\cite{Solovei2009} or senescent~\cite{Chandra2015,Zirkel2017} cells. When quenching from the UD phase into the DO region, which may model the mitosis-interphase transition, the system phase separates into competing epigenomic domains which slowly evolve into homogeneously marked systems. These transient states display epigenomic organisations that are reminiscent of typical cell nuclei~\cite{Cremer2015}. Yet, the long-time steady state lacks epigenetic state coexistence and is fully phase separated, so is qualitatively different from typical nuclear organisation. The metastable multidomain state can be stabilised though, by driving the system away from equilibrium as detailed below.

\begin{figure}[!t]
\includegraphics[width=0.45\textwidth]{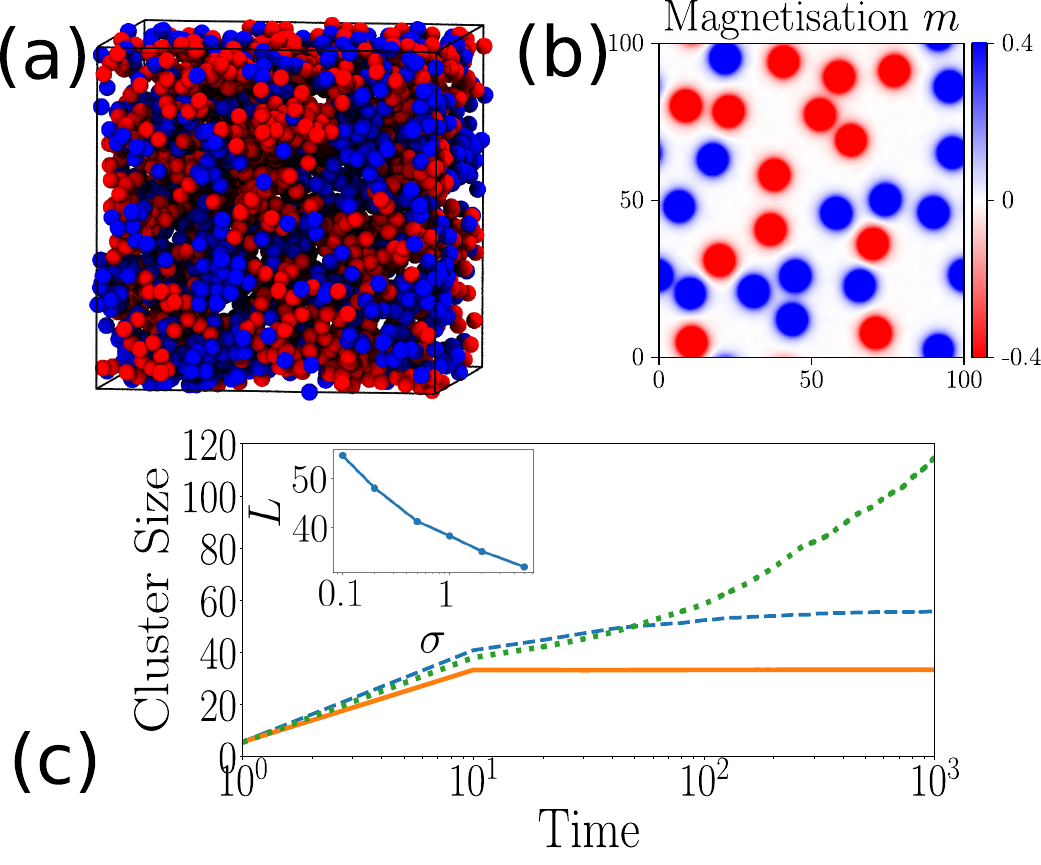}
\caption{\textbf{Non-Equilibrium Switching Drives Arrested Phase Separated Epigenomic Domains.} (a) Snapshot from a BD simulation of a melt of polymers with switching rate $\kappa=10^{-4} \tau_B$ ($\tau_B$ is the diffusion time of a monomer), monomer density $\rho=0.8\sigma^{-3}$ and $\varepsilon/k_BT_L=0.9$ (see SI for details). (b) Snapshot of a steady state configuration obtained evolving Eqs.~\eqref{eq:modelC+switch} with parameters $\Gamma_m=\Gamma_n=k_m=k_n=1$, $\chi = 6$, $n_0=0.5$, $\sigma_{a} =\sigma_i = 5$. (c) Evolution of typical epigenetic domain size as a function of time and for different switching rates. These figures show that the ordering dynamics is arrested and domains with well-defined (self-limiting) size are formed when epigenetic switching is included in the model. 
}
\vspace*{-0.5 cm}
\label{fig:Switching}
\end{figure}

\paragraph{Non-Equilibrium Epigenomic Organisation}
We now propose a non-equilibrium model for epigenomic organisation that can be derived starting from the free energy in Eq.~\eqref{eq:FMelt}. We consider its ``model C'' equations~\cite{ChaikinLubensky,Kockelkoren2002} and add two kinetic terms that dynamically convert the chromosomal density field from an ``active'' state, which can be biochemically marked ($n_a$)
to an ``inactive'' one that is refractory to biochemical modification ($n_i$), and vice versa. This switch is inspired by the process of ATP-dependent chromatin remodelling which changes local fibre structure and is coupled to histone modification~\cite{Alberts2014}. Note that now it is only the sum of the two density fields needs to be conserved at all times, i.e. $n_i+ n_a\equiv n_0$. The modified equations read
\begin{align}
&  \dot{m} = \Gamma_m \left(2 \chi m n_a -2 a m - 4 b m^3\right)+ \kappa_m \nabla^2 m  \notag \\
&  \dot{n}_a = \Gamma_n \nabla^2 \left(2 c n_a + 3 d n_a^2 - \chi m^2\right) -\kappa_n \nabla^4 n_a  + \sigma_{a} n_i - \sigma_{i} n_a  \notag \\
& \dot{n}_i = \Gamma_n \nabla^2 \left(2 c n_i + 3 d n_i^2\right) -\kappa_n \nabla^4 n_i-\sigma_{a} n_i + \sigma_{i} n_a
\label{eq:modelC+switch}
\end{align}
where the parameters $\sigma_{a/i}$ describe the rates at which chromatin is activated/inactivated. In general we will consider $\sigma_i \neq \sigma_a$ (see SI).  We numerically evolve Eqs.~\eqref{eq:modelC+switch} starting from the UD phase. Importantly, we find that the presence of non equilibrium switching terms now lead to arrest of both density phase separation and epigenetic ordering~\cite{Brackley2017biophysj}.  The system stabilises into coexisting domains with high local density and non-zero epigenetic signature separated by regions with low active density (see Fig.~\ref{fig:Switching}). 
Large-scale BD simulations of magnetic polymer melts in which beads are switched from a passive ``non-magnetisable'' state to an active ``magnetisable'' one at rate $\kappa$ confirm this phenomenology (see Fig.~\ref{fig:Switching}, and SI). 

\paragraph{Conclusions} We have proposed and solved models of magnetic polymers that can be used to describe the coupling between  epigenetic patterns and genome organisation both at the single chromosome and at the whole nucleus scale. 

For a single chromosome, our magnetic polymer model can be solved at the mean field level~\cite{Garel1988} and displays three possible phases in steady state. The phase diagram is in agreement with that found from BD simulations~\cite{Michieletto2016prx}, and the dynamics of the model generically entails uncontrolled spreading of the dominant epigenetic mark, which is reminiscent of epigenetic silencing dynamics {\it in vivo}~\cite{Pinter2012}. 
At the whole nucleus scale, we consider a Landau free energy density to describe the coupling between epigenetic states and chromosomal density. By combining dynamical mean field theory based on this free energy and direct BD simulations, we find that the model now leads to growth of many epigenetic domains with different marks, as found experimentally. In equilibrium one epigenetic domain eventually takes over the whole nucleus by spontaneous symmetry breaking. Unlimited spreading can though be contrasted by a non-equilibrium switching mechanism motivated by the phenomenon of ATP-dependent chromatin remodelling where each genomic segment can switch between a state in which it can be epigenetically marked and an inert one in which it cannot. 

Our magnetic polymer model for epigenomic ordering can be extended in a number of ways. One is by introducing genomic bookmarking to seed domain formation~\cite{Michieletto2018nar}. Another interesting avenue to explore would be to pursue a spin-glass model~\cite{EdwardsSpinGlass} instead of a Potts model for the underlying polymeric ordering. In this case, the rough free energy landscape of spin-glasses~\cite{Newman1999} might provide another avenue to stabilise a genome with micro-phase separated epigenetic domains.  

We thank the European Research Council (ERC CoG 648050 THREEDCELLPHYSICS) for funding.  

\bibliographystyle{apsrev4-1}
\bibliography{Epigenetics,Epigenetics2}

\begin{thebibliography}{64}
\bibitem [{not()}]{note1}%
  \BibitemOpen
  \href@noop {} {}\bibinfo {note} {Note that here we use loosely the word
  ``epigenetic'' to denote coarse-grained and monomeric states characterised by
  some biochemical histone modifications. In the current biology literature,
  biochemical marks or other features are said to be epigenetic only when they
  are stably inherited (at the single cell or whole organism
  level).}\BibitemShut {Stop}%
\end{thebibliography}

\newpage
\phantom{a}
\newpage
\appendix

\section{Supplementary Information}

\section{Single Chromosome Model}

Here we obtain the mean field approximation presented in the main text to describe the thermodynamics of 
a single chromosome fiber with epigenetic marks.

Following Ref.~\cite{Garel1999a}, we describe the chromosome fiber as a $N$-steps self-avoiding walk (SAW) 
on a lattice with coordination number $z$. Each vertex of the walk carries an epigenetic state $q$ that can 
assume three possible values ($q\in \{-1,0,1\}$).

Any pair of neighbouring (but non consecutive) vertices interact with each other via a contact potential that depends on their $q$-value.  More precisely, if the $i$-th and the $j$-th vertices are nearest neighbours on the lattice, their contact energy $J(q_i, q_j )$ is
\begin{small}
\begin{equation}\label{eq:contactenergy}
 J(q_i, q_j) = 
	      \begin{cases}
		      -\epsilon& \text{ if } q_i = q_j = \pm 1\\
		      0 & \text{ otherwise} 
	       \end{cases} \quad,
\end{equation}
\end{small}
with $\epsilon > 0$. Note that the mark $q=0$ does not contribute to this configurational energy and we will define it as a neutral mark. The equilibrium properties  of this system is described by the following partition function 
\begin{equation}\label{eq:partitionmarks}
 \mathcal{Z} = \sum_{\text{SAW}} \sum_{\{q\}} \exp\left[-\frac{\beta}{2} \sum_{i,j=1}^N \Delta_{\vec{r}_i ,\vec{r}_j} J(q_i, q_j)
 \right],
\end{equation}
where $1/\beta= k_B T$. The sums $\sum_{\text{SAW}}$ and $\sum_{\{q\}}$ run over the set of all $N$-steps SAWs and all the possible epigenetic states respectively. The matrix $\Delta_{\vec{r}_i \vec{r}_j}$ is the 
adjacency matrix associated to a given SAW and is given by
\begin{small}
\begin{equation}
 \Delta_{\vec{r}_i \vec{r}_j} = 
	\begin{cases}
	      1 & \text{if nearest neighbours} \\
	      0 & \text{otherwise}
	\end{cases} . 
\end{equation}
\end{small}
Notice that the partition function in Eq.~\eqref{eq:partitionmarks} presents a clear $\mathbb{Z}_2$ symmetry as $J(-q_i, -q_j) = J(q_i, q_j)$.

Since we are here interested in the critical properties of the system, we can restrict the phase space of the epigenetic variables, $q$, to the case where the abundance of the state $q=0$ is equal to the one of $q=-1$. With this restriction the system can be faithfully described by a two-valued spin variable $S=\{1,-\tfrac12\}$ where $S=1$ corresponds to the mark $q=1$, while the values $S=-\tfrac12$ has multiplicity 2 as it corresponds both to $q=0$ and $q=-1$ \cite{Wu1982}.

By using the spin variable $S$, Eq.~\eqref{eq:contactenergy} becomes
\begin{small}
\begin{equation}
 J(S_i, S_j) =
	      \begin{cases}
		      -\epsilon& \text{ if } S_i = S_j = 1\\
		      -\tfrac14\epsilon& \text{ if } S_i = S_j = -\tfrac12 \\
		      0 & \text{ otherwise} 
	       \end{cases} \quad.
\end{equation}
\end{small}
which can be re-written as
\begin{small}
\begin{equation}\label{eq:contactenergy2}
  J(S_i, S_j) = -\tfrac59\epsilon \left(S_i+\tfrac15\right)\left(S_j+\tfrac15\right) - \tfrac15\epsilon \, ,
\end{equation}
\end{small}
and the partition function in Eq.~\eqref{eq:partitionmarks} is then recast into
\begin{small}
\begin{equation} \label{eq:partitionspin}
\mathcal{Z} = \sum_{\text{SAW}} \sum_{\{S\}} \exp\left[\tfrac{\epsilon \beta }{2}\sum_{i,j=1}^N  
\Delta_{\vec{r}_i\vec{r}_j} \left(\tfrac59(S_i +\tfrac15)(S_j +\tfrac15)+ \tfrac15\right)\right]\,\,.
\end{equation}
\end{small}
Let us first evaluate, at a fixed $\gamma \in$ SAW,  the term:
\begin{small}
\begin{equation}\label{eq:firstterm}
\sum_{\{S\}} \exp\left[\frac{5\epsilon \beta }{18}\sum_{i,j=1}^N (S_i +\tfrac15)\Delta_{\vec{r}_i\vec{r}_j} (S_j +\tfrac15)\right] \quad. 
\end{equation}
\end{small}
By using an Hubbard-Stratonovich transformation Eq.~\eqref{eq:firstterm} becomes 
\begin{small}
\begin{equation*}
\int \mathrm{d} \vec{\phi} \exp\left[- \tfrac{9}{10 \epsilon \beta} \sum_{i,j=1}^N \phi_i \Delta_{\vec{r}_i\vec{r}_j}^{-1} \phi_j 
+\sum_{i=1}^N \log\left(\sum_{\{S_i\}} e^{\phi_i(S_i+\tfrac15)}\right)\right]
\end{equation*}
\end{small}
where $\mathrm{d}\vec{\phi} = \prod_{i=1}^N \mathrm{d} \phi_i$. By summing over all possible spin configurations we get
\begin{small}
\begin{equation}
\int \mathrm{d} \vec{\phi} \exp\left[- \tfrac{9}{10 \epsilon \beta} \sum_{i,j=1}^N \phi_i \Delta_{\vec{r}_i\vec{r}_j}^{-1} \phi_j 
+\sum_{i=1}^N \log\left(e^{\tfrac65\phi_i}+2e^{-\tfrac{3}{10}\phi_i}\right)\right].
\end{equation}
\end{small}
This integral can be evaluated through an homogeneous saddle point approximation and by assuming the translational invariance of the field $\phi$. This gives
\begin{small}
\begin{equation}\label{eq:hubb1}
\exp\left[- \tfrac{9}{10 \epsilon \beta} \phi^2 \sum_{i,j=1}^N  \Delta_{\vec{r}_i\vec{r}_j}^{-1} + N  \log\left(e^{\tfrac65\phi}+2e^{-\tfrac{3}{10}\phi}\right)\right]\quad.
\end{equation}
\end{small}
In general, the term $ \sum_{i,j=1}^N  \Delta_{\vec{r}_i\vec{r}_j}^{-1}$, depends on the given SAW and
it is not easy to compute. However, it can be estimated if we restrict the set of SAWs to the ones that are almost space filling, i.e. ones that can be approximated as Hamiltonian walks~\cite{Garel1999a}. 

An Hamiltonian walk is a path that visits each vertex of a lattice embedded in a volume $V$ exactly once and have been used to study equilibrium properties of highly compact polymers~\cite{Duplantier1987,Orland1985}. For an Hamiltonian walk, the adjacency matrix of the SAW $\Delta$ takes the same form of the adjacency matrix of the underlying lattice and it is characterised by the coordination number $z$. Hence, $\sum_{i,j=1}^N  \Delta_{\vec{r}_i\vec{r}_j}^{-1} = \frac{N}{z}$. Here, we consider $N$-steps configurations that, similarly to Hamiltonian walks, are contained in a volume $V$ but may in principle display a lower mean number of nearest neighbours, i.e. $\rho z$ instead of $z$. With this approximation 
\begin{equation}
 \sum_{i,j=1}^N  \Delta_{\vec{r}_i\vec{r}_j}^{-1} \approx \frac{N}{\rho z} \quad.
 \label{eq10}
\end{equation}
Notice that for generic SAWs with low $\rho$ values Eq.~\ref{eq10} is not exact but is an upper bound.

Finally, we  evaluate the last term in Eq.~\eqref{eq:partitionspin}, i.e.
\begin{small}
\begin{equation}
e^{F_{\text{SAW}}} = \sum_{\gamma \in \text{SAW}} 
\exp\left[\frac{\beta \epsilon}{10} \sum_{i,j=1}^N \restrdelta_{ij} \right]. \
\end{equation}
\end{small}
By following the approach described in Ref.~\cite{Nemirovsky1992} we can approximate $F_\text{SAW}$ as
\begin{equation}\label{eq:FSAW}
\frac{F_\text{SAW}}{T N}  \approx - \log\left(\frac{z}{e}\right)+ \frac{1-\rho}{\rho}\log(1-\rho) - \frac{\beta \epsilon z}{10 }\rho\quad.
\end{equation}

By collecting all the terms and taking $f = -\frac{T}{N}\log\mathcal{Z}$, we obtain the following mean-field free energy density
\begin{equation}\label{eq:freeenergysingle}
\begin{split}
\frac{f}{T} =&   - \log\left(\frac{z}{e}\right)+ \frac{1-\rho}{\rho}\log(1-\rho) - \frac{\alpha}{10 }\rho + \\
& +  \frac{9}{10\alpha} \frac{\phi^2}{\rho}  - \log\left(e^{\tfrac65\phi}+2e^{-\tfrac{3}{10}\phi}\right)
\end{split}\quad,
\end{equation}
where $\alpha \equiv \beta \epsilon z$.
The equilibrium properties of the model are then obtained by minimizing Eq.~\eqref{eq:freeenergysingle} with respect to both, magnetisation $\phi$ and density $\rho$. As stated in the main text, 
this mean field approximation gives two possible equilibrium phases. 
For large values of $\alpha$ we find a 
\emph{compact-ordered} phase (CO) where the chain is globular ($\rho\ne 0$) and nearly uniformly coloured $(\phi > 0)$.
For small values of $\alpha$ the system is instead in a \emph{swollen-disordered} phase (SD) where
the chain is extended in space ($\rho=0$) and heterogeneously coloured ($\phi=0$). At the transition point ($\alpha_c\simeq 3.96$) we observe a discontinuous jump of the parameters 
$\rho$ and $\phi$, proving the existence of a first order transition between the two phases~\cite{Michieletto2016,Michieletto2017}.

\section{Genome-wide Model}

Here, we discuss the model we introduced in the main text to describe the equilibrium properties of epigenomic organisation at the scale of the full genome. By assuming that chromosomes fill a fixed volume $V$, we can define a conserved mean density $n_0 \equiv N/V$,
where $N$ is the total length of the genome. The equilibrium properties can be described by the following free-energy density
\begin{equation}\label{eq:FMelt}
 \beta f = a m^2 + b m^4 + c n^2 + d n^3 -\chi m^2 n \,,
\end{equation}
where the fields $n(\vec{x}, t)$ and $m(\vec{x},t)$ correspond to the local chromatin density distribution and the average epigenetic marks (or magnetisation) respectively.  The phenomenological parameters of the uncoupled system are constant and set to be $a> 0$, $b> 0$, $c> 0$, $d > 0$. The parameter $\chi>0$, governing the coupling between the epigenetic profile and the chromatin organisation, is temperature dependent. Since $V$ is fixed, the local density $n$ obeys the following constraint:
\begin{equation}\label{eq:constraint}
 n_0 = \frac{1}{V}\int_V n(\vec{x}, t) \,\mathrm{d} \vec{x}\qquad \forall t \,.
\end{equation}

The equilibrium properties are found by minimizing the functional $\mathcal{F}= \int_V f(\vec{x})\,\mathrm{d}\vec{x}$ with the constraint in Eq.~\eqref{eq:constraint}, i.e. constant $n_0$.
This is equivalent to find the minima of the functional $\mathcal{G} = \mathcal{F} -  \mu \int_V \left[n(\vec{x},t)-n_0\right]\,\mathrm{d}\vec{x}$ i.e. to solve
the set of equations:
\begin{equation} \label{eq:minimizingcomplete}
\left\{ \begin{array}{l}
	  \delta f\left[m(\vec{x}),n(\vec{x})\right] / \delta m = 0 \\
	  \delta f\left[m(\vec{x}),n(\vec{x})\right] / \delta n= \mu \\
	  \frac1V \int_V n(\vec{x}) \,\mathrm{d} \vec x = n_0  
	\end{array}
\right.	         \ \ ,
\end{equation}
where $f\left[m(\vec{x}),n(\vec{x})\right]$ denotes the free energy functional and $\delta f / \delta m$ the functional derivative.

By finding the solution to the first equation, i.e. 
\begin{equation} \label{eq:mequilibrium}
\ateq{m}\left[n(\vec{x},t)\right] = \begin{cases} 0 & \text{if } n(\vec{x},t) \leq \frac{a}{\chi} \\ 
				    \pm\frac{\sqrt{\chi \,n(\vec{x},t) - a}}{\sqrt{2b}} & \text{if } n(\vec{x},t) > \frac{a}{\chi} 
			\end{cases} \ \ ,
\end{equation}
we restrict the problem to the effective free-energy density $f^\star \equiv f^\star\left[n(\vec{x},t)\right] \equiv f\left[\ateq{m}(\vec{x},t),n(\vec{x},t) \right]$ that
depends only on the conserved field $n$ and reads:
\begin{small}
\begin{equation}
f^\star = \begin{cases} 
					c n^2 + d n^3 & \text{if }  n(\vec{x},t) \leq \frac{a}{\chi}  \\
					-\frac{a^2}{4 b}+\frac{a   \chi }{2 b} n+ \left(c-\frac{\chi ^2}{4b}\right)n^2+d n^3 & \text{otherwise}
   \end{cases}\quad .
\end{equation}
\end{small}
This procedure simplifies Eqs.~\eqref{eq:minimizingcomplete} to the set of equations
\begin{equation} \label{eq:minimizingdensity}
\left\{ \begin{array}{l}
	  \delta f^\star / \delta n= \mu \\
	  \frac1V \int_V n(\vec{x}) \,\mathrm{d} \vec x = n_0  
	\end{array}
\right.
\end{equation}
which is satisfied by the trivial uniform solution $n(\vec{x}) \equiv n_0$.  In the non-trivial solution of these equations, instead, we find that in the system there is a cohexistence between two density phases $n(\vec{x}) = n_-$ and $n(\vec{x}) = n_+$, have the same pressure $P =f^\star - n \derfunc{f^\star}{n}$ and chemical potentials $\mu$, and are found via the so called common tangent construction~\cite{ChaikinLubensky}.

%

 
Finally, the (spinodal) region in which the homogeneous solution $n(\vec{x}) = n_0$ is unstable is characterised by $\left. \derfunc{^2f^\star}{n^2}\right|_{n(\vec{x}) = n_0} < 0$, which leads to
\begin{equation}\label{eq:spinodal}
\frac{a}{\chi} < n_0 < \frac{\chi ^2-4 b c}{12 b d} \ \qquad.
\end{equation}
Inside this region of values the homogeneous solution is linearly unstable and the system spontaneously demixes into low density ($n_-$) and high density ($n_+$) phases.

By applying this procedure to the free-energy in Eq.\eqref{eq:FMelt} we obtain the equilibrium phase-diagram as a function of the coupling parameter $\chi$ and genome density $n_0$ (see main text and Fig.~2).

\subsection{Nature of the Phase Transitions}
We now discuss the nature of the lines of phase transitions found in the equilibrium phase diagram: 

First, from Eq.~\eqref{eq:mequilibrium} one can notice that the order parameter $m$ goes continuously to zero. This strongly suggests that the transition from UD to UO is second order. Second, if a system is driven from the homogeneous phase, where $n(\vec{x}) = n_0\,\,\forall\vec{x}$, to a region in which this solution becomes unstable, then it must cross a binodal line. At this point the pressure is the critical one $P^\star$ and we find that either
\[\lim_{\{\chi, n_0\}\to P^\star} n_- = n_0 \quad\text{ or }\quad\lim_{\{\chi, n_0\}\to P^\star} n_+ = n_0\ \ .\] 
Similarly, if the system is driven from one demixed region (e.g. PDO with $0 < n_-^\prime < n_0 <  n_+^\prime$) to another (e.g. DO with $0=n_-<n_0 < n_+$), then the system must cross another point $P^\star$ where
\[\lim_{\{\chi, n_0\}\to P^\star} n_- = n_-^\prime \quad\text{ and }\quad\lim_{\{\chi, n_0\}\to P^\star} n_+ = n_+^\prime\ \ .\]

In light of this, and of the fact that the magnetic order parameter is continuous, we can conclude that every transition line in the phase diagram is continuous. Below we will focus in more detail on the transition from the homogeneous to the demixed phase, but a similar argument can be used for phase transitions between demixed phases.

A homogeneous phase displays $\mathcal{F}^{(eq)}_H/V = f^\star(n_0; \chi)$ while in a demixed one  $\mathcal{F}^{(eq)}_D/V = \alpha f^\star(n_-; \chi) + (1-\alpha) f^\star(n_+; \chi)$, where $0 \leq\alpha\leq1$ is such that $\alpha n_-+(1-\alpha)n_+ = n_0$. By looking at the first derivatives of the free energy in Eq.~\eqref{eq:FMelt}, computed at the equilibrium, and using the above conditions we get 
\[\derpar{\alpha (n_+-n_-)}{\chi}  = \alpha \derpar{n_-}{\chi} + (1-\alpha) \derpar{n_+}{\chi} \ \ \ .\]
This equality,  together with the common tangent construction which gives the constraints $\left.\derpar{f}{n}\right|_{n=n_-} = \left.\derpar{f}{n}\right|_{n=n_+} $
and $f(n_-)-n_-\left.\derpar{f}{n}\right|_{n=n_-} = f(n_+)-n_+\left.\derpar{f}{n}\right|_{n=n_+}$, leads to:
\begin{equation}
\frac1V\derpar{\mathcal{F}^{(eq)}_D}{\chi} = \alpha \left.\derpar{f}{\chi}\right|_{n=n_-} + ( 1-\alpha) \left.\derpar{f}{\chi}\right|_{n=n_+}\qquad.
\end{equation}
Since the order parameter $n$ is continuous, if we drive the system from the homogeneous phase, we expect that either $\alpha \to 0$, $n_+\to n_0$ or $\alpha \to 1$, $n_-\to n_0$. Hence  
\begin{equation}
\lim_{\{\chi, n\}\to P^\star}\frac1V\derpar{\mathcal{F}^{(eq)}_D}{\chi} = \left.\derpar{f}{\chi}\right|_{n=n_0} \equiv \frac{1}{V} \derpar{\mathcal{F}^{(eq)}_H}{\chi}.
\end{equation}
Similarly, one can show that:
\begin{equation}
\lim_{\{\chi, n\}\to P^\star}\frac1V\derpar{\mathcal{F}^{(eq)}_D}{n_0} = \left.\derpar{f}{n}\right|_{n=n_0} \equiv \frac{1}{V} \derpar{\mathcal{F}^{(eq)}_H}{n_0}.
\end{equation}
Therefore,  as the system pass from an homogeneous phase, to a demixed one, the first derivatives of the free energy are continuous. 

\begin{figure*}[!t]
	\includegraphics[width=0.7\textwidth]{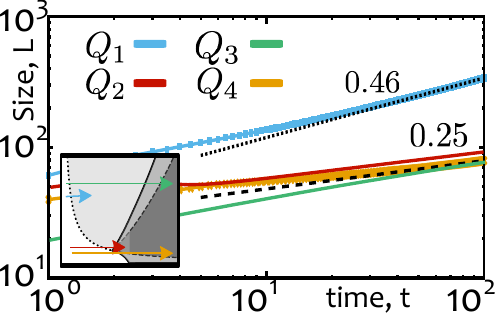} 
	\caption{\textbf{Growth of Epigenetic Domains.} Evolution of typical epigenetic domain size $L$ following a quench from the uniform disordered phase. The growth displays a power law that is compatible with Model A dynamics ($\alpha=1/2$) when the density field remains uniform (quench $Q_1$), but it is significantly slowed down when both fields are re-organised ($\alpha=1/4$). In the inset we schematically show the paths of the quenches in the phase diagram.  Here we evolved Eqs.~\eqref{eq:modelc} with fixed $a=b=c=d=1$ and $\Gamma_m = \Gamma_n = \kappa_m = \kappa_n = 1$.} 
	\label{fig:crit_exp}
	
\end{figure*}

\section{Dynamical Scaling}

Here we characterize the dynamical evolution of the system when it is quenched
from a point within the UD phase into one within either the UO or the DO phase. 
This analysis may provide insights into the dynamics of the genome-wide spatial re-organisation and epigenetic recolouring, for example, at the beginning of interphase.
In this model the density is a conserved order parameter while the magnetisation needs not be conserved. Hence, the dynamics of the system can be described by ``model C''~\cite{ChaikinLubensky} equations:
\begin{equation}
\left\{\begin{array}{rcl}
\derpar{m}{t} &=& -\Gamma_m \frac{\delta f}{\delta m} + D_m \nabla^2 m \nonumber \\
\partial_t n &=& \Gamma_n \nabla^2  \frac{\delta f}{\delta m} -D_n \nabla^4 n \, ,
\end{array}\right. 
\label{eq:modelcBasic}
\end{equation} 
by using the free energy in Eq.~\eqref{eq:FMelt} these become
\begin{eqnarray}
\partial_t m &=& \Gamma_m \left(2 \chi m n -2 a m - 4 b m^3 \right)+ D_m \nabla^2 m \nonumber \\
\partial_t n &=& \Gamma_n \nabla^2 \left(2 c \rho + 3 d \rho^2 - \chi m^2\right) -D_n \nabla^4 n \, .
\label{eq:modelc}
\end{eqnarray} 
We numerically solve Eqs.~\eqref{eq:modelc} and monitor the time evolution of the density and magnetisation fields during several possible quenching trajectories in the phase space.
We start from the UD phase and perform four representative quenches: 
$Q_1$: Uniform Disordered $\rightarrow$ Uniform Ordered; $Q_2$: Uniform Disordered $\rightarrow$ Partially-Demixed Ordered; $Q_3$: Uniform Disordered $\rightarrow$ Demixed Ordered (large $n_0$);$Q_4$: Uniform Disordered $\rightarrow$ Demixed Ordered (small $n_0$) (see inset of Fig.~\ref{fig:crit_exp}). 

Following $Q_1$, we observe that the density remains uniform while the epigenetic field coarsens into clusters of coherent colours which slowly evolve into one system-spanning domain through spontaneous symmetry breaking (see {\bf movie M1}). The scaling of the typical epigenetic domain size grows as $L(t) \sim t^{\alpha}$ where $\alpha = 0.46$ is compatible with Model A dynamics~\cite{Kockelkoren2002} (see Fig.~\ref{fig:crit_exp}). This is expected since the density field remains uniform. 

\begin{figure*}[!t]
 \centering
 \includegraphics[width=\textwidth]{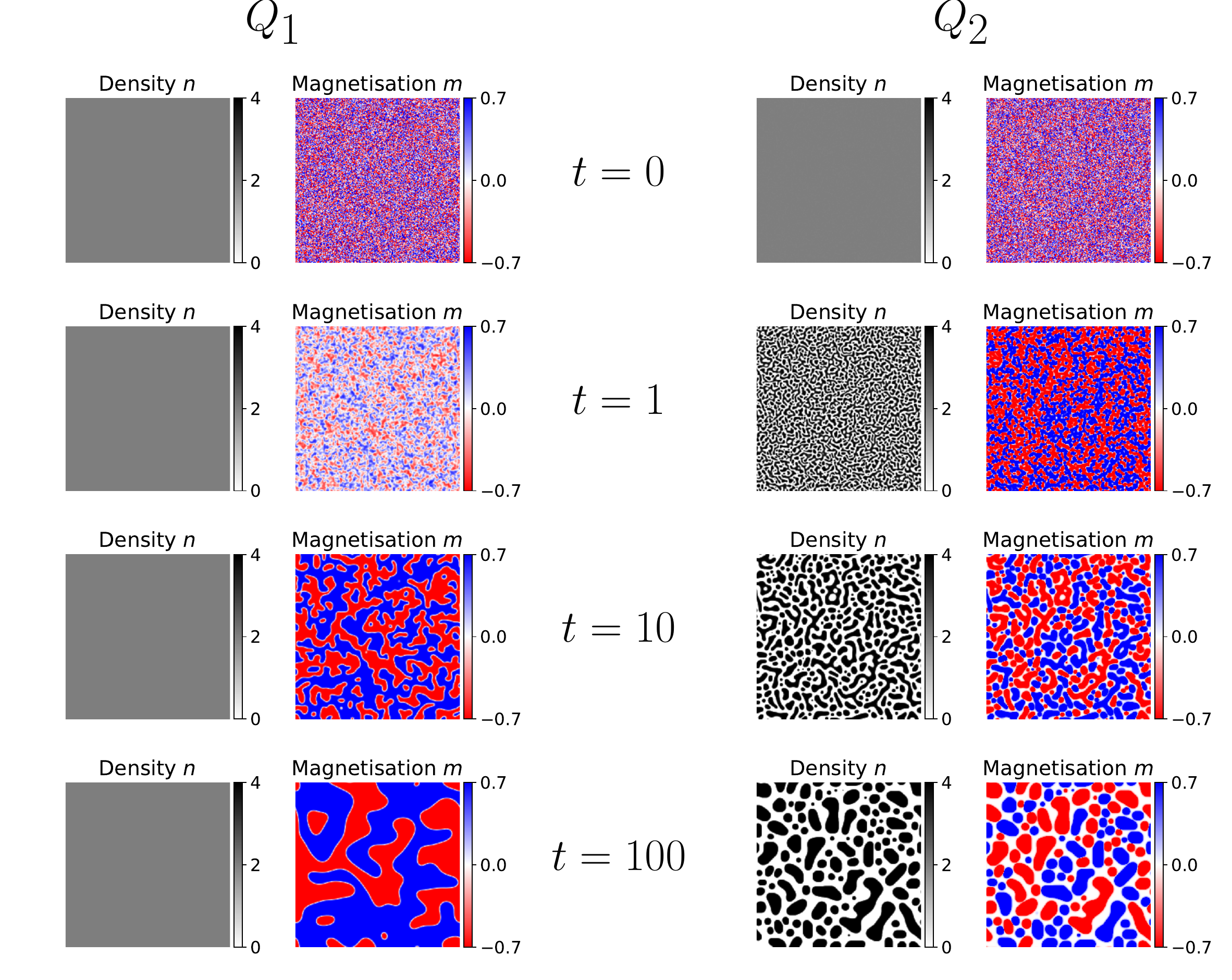}
 \caption{Representative snapshots of the system at different times of its evolution for two quenches. On the left-hand side, the quench $Q_1$ brings the system from an UD phase to the UO phase: while the density field remains homogeneous, the magnetisation shows the formation of clusters which slowly evolve into a uniformly coloured system. Here the dynamics is akin to the Model A one. On the right-hand side, $Q_3$ brings the system from the UD phase to the DO phase: in this case both magnetisation and density fields needs to re-organize and the system evolves on slower time-scales. 
 }\label{fig:snapshots}
\end{figure*}

We also observe that the other three quenches evolve on slower timescales as both fields need to be re-organised since we drive a transition from a homogeneous system to a demixed one (see Fig.~\ref{fig:snapshots}). Specifically, for quenches $Q_2$, $Q_3$, and $Q_4$, $L_m(t) \sim t^{\beta}$ with $\beta \simeq 0.25$ in agreement with previous results on Model C dynamics~\cite{Kockelkoren2002} (see \textbf{Movie M2}, \textbf{M3}, \textbf{M4}).

\begin{figure*}[t!]
	\includegraphics[width=\textwidth]{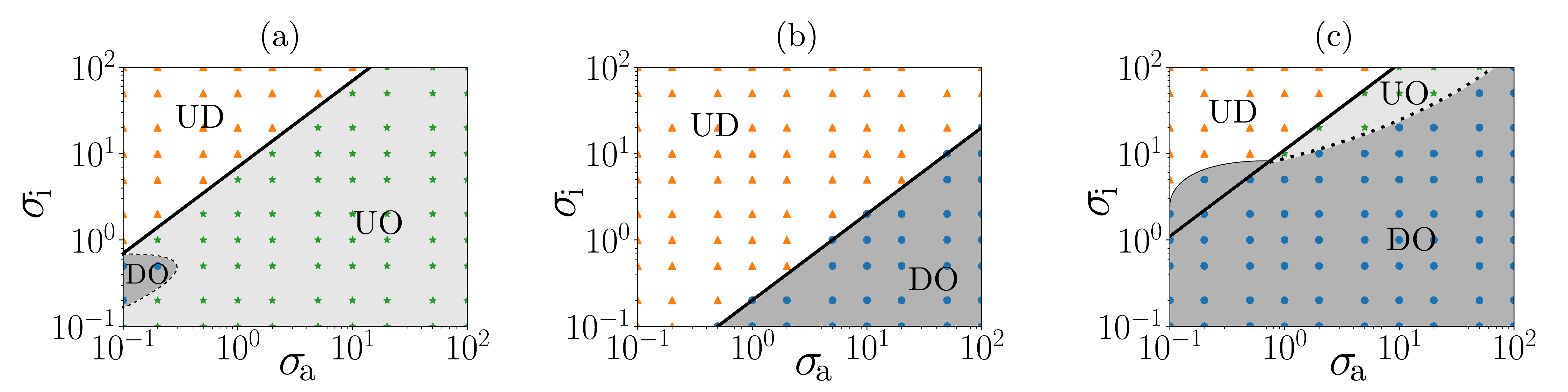}
	\caption{Non-equilibrium phase diagrams obtained by integrating numerically Eqs.~\eqref{eq:modelC+switch} on a 2-dimensional grid of side $L=100\times 100$,cwith $a=b=c=d=1$, $\Gamma_m=\Gamma_n =1$, $\kappa_m=\kappa_n=1$. All the figures are obtained at fixed values of $\chi$ and $n_0$. We employed  (a) $\chi=4$, $n_0=2$ (quench into UO); (b) $\chi=6$, $n_0=0.2$ (quench into DO);  (c) $\chi=6$, $n_0=2$ (quench into DO).  The solid black line is found via Eq.~\eqref{ratioswitch} and should separate the ordered phases from the the disordered one (UD). In the figures, different symbols highlight different phases of the system as indicated by the names (UD, UO, DO). 
	}\label{fig:phasediagrams_switching}
\end{figure*}

\section{Non-Equilibrium Epigenetic Switching}

Here we present the details of the non-equilibrium model for genome organisation with epigenetic switching. As reported in the main text, the dynamical equations are
\begin{align}
&  \derpar{m  }{t} = \Gamma_m \left(2 \chi m n_a -2 a m - 4 b m^3\right)+ \kappa_m \nabla^2 m  \notag \\
&  \derpar{n_a}{t} = \Gamma_n \nabla^2 \left(2 c n_a + 3 d n_a^2 - \chi m^2\right) -\kappa_n \nabla^4 n_a  + \sigma_{a} n_i - \sigma_{i} n_a  \notag \\
&  \derpar{n_i}{t} = \Gamma_n \nabla^2 \left(2 c n_i + 3 d n_i^2\right) -\kappa_n \nabla^4 n_i-\sigma_{a} n_i + \sigma_{i} n_a\, .
\label{eq:modelC+switch}
\end{align}
Eqs.~\eqref{eq:modelC+switch} describe the dynamics of a ``model C''~\cite{ChaikinLubensky} with two additional kinetic terms
that dynamically convert the density fields from one that can be epigenetically marked (or active, $n_a$)
to the one that is uncoupled from the epigenetic field (or inactive, $n_i$).  As discussed in the main text, these terms may effectively account for the non-equilibrium action of so-called chromatin remodelling complexes~\cite{Alberts2014} that render a local genomic region available for, or refractory to, epigenetic marking at a certain time.  The amplitudes of $\sigma_{a/i}$ describe the rates at which the density fields $(n_a,n_i)$ are activated/inactivated, i.e. the rates at which chromatin remodelling factors act on the genome.

One should notice that in this case the total density $n_a+n_i$ must be conserved, i.e.
\begin{equation}\label{eq:constraint_switching}
 \frac{1}{V}\int_V \left[n_a(\vec{x},t)+n_i(\vec{x},t)\right]d\vec{x} = n_0 \qquad \qquad \forall t \, ,
\end{equation}
whereas $n_a$ and $n_i$ need not to be individually conserved.
Nevertheless, since $\partial (n_a + n_i)/\partial t$ can be written as the divergence of a certain quantity, 
equation Eq.~\eqref{eq:constraint_switching} is always satisfied.  

We also mention that by imposing a free energy of the form:
\[
 f = a m^2 + b m^4 + c n_a^2 + d n_a^3- \chi m^2 n_a + c n_i^2 + d n_i^3 + \mathcal{G}
\]
where $\mathcal{G}(\vec{x}, t)$ is a function such that:
\[\nabla^2 \derpar{\mathcal{G}}{n_i} = \sigma_a n_a -\sigma_i n_i = - \nabla^2 \derpar{\mathcal{G}}{n_a},\ \ \ \]
then Eqs.~\eqref{eq:modelC+switch} can be derived from an effective free energy only if $\sigma_a=\sigma_i$. In this case $\mathcal{G}$ takes the form
\[\mathcal{G}(\vec{x}) = \frac{\sigma}{2} \int_V G(\vec{x}-\vec{y})\left(n_i(\vec{y}) - n_a(\vec{y})\right)^2 \,\,\mathrm{d}\vec{y},\]
where $G(\vec{x})$ is the Green function that solves the equation $\nabla^2 G(\vec{x})= \delta(\vec{x})$ and it depends on the system dimension. Note that, if the switching rates are equal, $\sigma_i=\sigma_a$, then the dynamical equations of the system can be understood as underlying an effective free energy, thus entailing that the system is in equilibrium. On the other hand, the general condition that $\sigma_a\neq \sigma_i$, entails that Eqs.~\eqref{eq:modelC+switch} describe a purely non-equilibrium system.

\subsection{Steady States of the Switching Model} 
We now study the dynamics and the steady states of the model described by Eqs.~\eqref {eq:modelC+switch} varying the values of $\sigma_i$ and $\sigma_a$. We keep the phase diagram of the system (Fig.~2 of main text) as a reference and fix the values of $n_0$ and $\chi$ such that 
a phase in the limit of negligible density of inactive marks, i.e. $\sigma_a \gg \sigma_i$ can be observed.

If we quench the system either into the Uniform Ordered or the Demixed Ordered phases, then by varying $\sigma_i$ and $\sigma_a$ leads to a new stationary state similar to the Partially Demixed Ordered phase, i.e. one characterised by weak variations of the total density $(n_{-} > 0)$ and denoted by a non-null magnetisation $m^2>0$ (see \textbf{movie MS1}, \textbf{movie MS2}, \textbf{movie MS3}, \textbf{movie MS4}). 

The non-equilibrium phase diagrams of the model at fixed $n_0$ and $\chi$ as a function of the two kinetic rates $\sigma_{a/i}$ are shown in Fig.~S\ref{fig:phasediagrams_switching}. 
In most of the cases, these pictures show that the  
ordered phases arise when the fraction $\sigma_i/\sigma_a$ is lower than a certain critical ratio $r$ which can be estimated as follows: 
in steady state, Eqs.~\eqref{eq:modelC+switch} predict a 
mean active density 
$$\braket{n_a} = \frac1V \int_V n_a \mathrm{d}\vec{x} \approx \dfrac{\sigma_a n_0}{\sigma_i+\sigma_a} \, .$$
On the other hand, in Eq.~\eqref{eq:spinodal} we have shown that the ordered states are stable only if the active density $\braket{n_a} >\frac{a}{\chi}$. Thus, one can conclude that the Ordered phases (Uniform or Demixed) are strongly favoured if
\begin{equation}\label{ratioswitch}
 \sigma_i < \left(1 - \frac{n_0 \chi}{a}\right)\sigma_a \, ,
\end{equation}
in very good quantitative agreement with the observations from the numerical evolution of the system (see Fig.~S\ref{fig:phasediagrams_switching} black lines).
 
As discussed in the main text, the Demixed Ordered phase observed in this model is very different from the one achieved in equilibrium. Indeed, here we observe an arrested coarsening of the epigenetic domains whose self-limiting size can be directly tuned by the kinetic parameters $\sigma_{a/i}$. We highlight that the concept of non-equilibrium switching has been applied in the literature to show that clusters of proteins can display an arrested coarsening and continuous recycling with the soluble pool~\cite{Brackley2017biophysj} but never applied to the dynamics of epigenetic marks.

\section{Brownian Dynamics Simulations of Annealed Copolymers} \label{sec:Brownian}
Here we describe the model employed for performing Brownian Dynamics  (BD) simulations of chromosomes with dynamic epigenetic marks.

Chromosomes are modelled using semi-flexible bead-spring chains~\cite{Kremer1990} as successfully done in the literature~\cite{Rosa2008,Mirny2011}. Each bead is marked with an epigenetic state $q=\{-1,0,1\}$ and the dynamics of the chains are described by a set of Langevin equations at the temperature $T_L$. After evolving the dynamics of a $N$-beads long chain for a certain time $\tau_R$, we evolve the colour of the beads using a number $N$ of Metropolis moves at the temperature $T_R$. This process is repeated several times, until the system achieves a steady state.
 
The Hamiltonian that describes the system is of the form
 \begin{equation}\label{eq:Hamiltonian_Brownian}
    H = \sum_{i=1}^M \dfrac{m}{2} \left(\frac{\mathrm{d} \vec{r}_i}{\mathrm{d}t}\right)^2 + U\left(\set{\vec{r}},\set{q}\right) \text{ ,}
 \end{equation}
where the first term is the kinetic one, while the second is a general interaction term between the beads. In our case, we model the interactions as follows:
 \begin{equation}\label{eq:TotalInteraction}
 U = \mathrm{U}_{\text{H}}\left(\vec{r}\right)+ \mathrm{U}_{\text{K}}\left(\vec{r}\right) + \mathrm{U}_{\text{LJ}}\left(\vec{r},q\right) \text{ ,}
\end{equation} 
 where:
 \begin{enumerate}
 \item $\mathrm{U}_{\text{K}}$ is a Kratky-Porod term which models the stiffness of the chain:
 \begin{small}
 \begin{equation}\label{eq:Kratky_Potts}
\frac{\mathrm{U}_{\text{K}}\left(\set{\vec{r}}\right)}{k_B T_L}  = \frac{\ell_P}{\sigma}\sum_{i=1}^{M-2}\left( 1 - \frac{ \vec{u}_i \cdot \vec{u}_{i+1}}{\norm{\vec{u}_i}\norm{\vec{u}_{i+1}}}\right) \text{ ,}
 \end{equation}
 \end{small}
 where $\vec{u}_j \equiv \vec{r}_{j+1}-\vec{r}_j$ and $\ell_P$ is identified with the persistence length of the chain, here set to $\ell_P = 3 \sigma \simeq 90 $ nm to match that of chromatin~\cite{Dekker2002}.
 
 \item $\mathrm{U}_{\text{LJ}}$ describes excluded volume interactions:
 \begin{small}
 \begin{equation}\label{eq:TotalLJ_Potts}
 \mathrm{U}_{\text{LJ}}\left(\set{\vec{r}}, \set{q} \right) = \sum_{j>i}  U_{\text{LJ}}(\norm{\vec{r}_i - \vec{r}_j}; q_i, q_j) \text{ ,}
  \end{equation}
 \end{small}
 with $U_{\text{LJ}}$ being a truncated and shifted Lennard-Jones potential, i.e. 
 \begin{small}
 \begin{equation} \label{eq:LJ}
 \begin{split}
  \frac{U_{\text{LJ}}(r;q_i, q_j)}{k_B T_L} = &\frac{ 4  }{\mathcal{N}}\frac{\varepsilon(q_i,q_j)}{k_B T_L}\left[ \left(\frac{\sigma}{r}\right)^{12}
  -  \left(\frac{\sigma}{r}\right)^6 +\right.\\&
  \left.-\vphantom{\frac{ 4  }{\mathcal{N}}} U_0(r_c(q_i,q_j)) \right] 
  \Theta\left(r-r_c(q_i,q_j)\right) \text{ ,}
  \end{split}  
 \end{equation}
 \end{small}
 where $\Theta$ is the Heaviside step function, $U_0$ is an auxiliary function which ensures that $U_{\text{LJ}}(r_c(q_i,q_j);q_i, q_j) \equiv 0$ and the cutoff $r_c(q_i,q_j)$ is $q-$dependent. 
 In particular we set:
 \begin{enumerate}
  \item $r_c(q_i, q_j) = 2^{1/6}\sigma$ if $q_i \neq q_j$ or $q_i=q_j = 0$, modeling only steric interaction between beads with different epigenetic marks or unmarked ($q= 3$);
  \item $r_c(1,1) = r_c(-1,-1) =  1.8\sigma$, modeling the effective attractive interaction between beads with the same epigenetic
  marks mediated by the ``readers'' enzymes~\cite{Michieletto2016}.
 \end{enumerate}
 Finally, the free parameter $\varepsilon(q_i, q_j)$ is:
 \begin{equation}
  \frac{\varepsilon(q_i,q_j)}{k_B T_L} = 
			  \left\{\begin{array}{ll}
				  \frac{\epsilon}{k_B T_L}& \text{if } q_j=q_j =\pm 1\\
                                 1 & \text{otherwise} 
                                \end{array}
			  \right.
 \end{equation}
 and $\mathcal N$ is a parameter which ensures that the minimum of the attractive part is $-\frac{\epsilon}{k_B T_L}$.
 \item $\mathrm{U}_{\text{H}}$ describes the connection between consecutive beads along the chain:
 \begin{small}
 \begin{equation}\label{eq:Harmonic_Potts}
  \frac{U_{\text{H}}(\set{\vec{r}})}{k_B T_L} =\sum_{i=1}^{M-1} \frac{k_h}{2 k_B T_L} \left( \left\| \vec{r}_i - \vec{r}_{i+1} \right\| - r_0\right)^2 \, ,
 \end{equation}
 where $k_h$ models the connectivity strength and it is set ot $k_H = 200 \epsilon$.
 \end{small}
 
\end{enumerate}
 
We then use these potentials to evolve the equations of motion for each bead in the system using a fixed-volume and constant-temperature molecular dynamics (MD) simulations (NVT ensemble). The simulations are run within the LAMMPS engine~\cite{Plimpton1995} and the equations of motion are integrated using a velocity Verlet algorithm, in which all beads are weakly coupled to a Langevin heat bath with friction $\gamma = \tau_B^{-1}$ where $\tau_B = 3\pi \eta \sigma^3/k_BT$ is the self-diffusion (Brownian) time of a bead of size $\sigma$ moving in a solution with viscosity $\eta$ (which we consider water, i.e. $\eta= 1 cP$, for the mapping to real units). Finally, the integration time step is set to $\Delta \tau = 0.01 \, \tau_B$.
 
As mentioned before, ``recolouring'' steps are performed every $\tau_R=100 \tau_B$ and in each step we attempt a number of moves equal to the number of beads in the system. In each move, we randomly select a bead and randomly change its colour to a different one. If the move lowers the energy of the system we accept it, otherwise we assign an acceptance probability $p=e^{-\Delta E/k_BT_R}$ where $\Delta E$ is the change in system energy after and before the move. 

In this scheme, it is straightforward to implement non-equilibrium switching by defining a fourth bead type (or $q=2$) which does not participate to the recolouring dynamics, i.e. beads bearing $q=2$ are excluded from the recolouring moves. Then, at rate $\sigma_i$, beads bearing $q=\{-1,0,1\}$ are randomly converted into $q=2$ and viceversa at rate $\sigma_a$. 
 
\subsection{Single Chromosomes}
We employ single chromosome BD simulations of this model to confirm the results obtained through our continuum model in Fig.~1 of the main text.
In the equilibrium case ($T_L=T_R$) the main parameter that is varied to confirm the phase diagram reported in Fig.~1a is $\alpha=\varepsilon/k_BT_L$. In the non-equilibrium case, we break detailed balance and independently vary $T_L$ and $T_R$ while maintaining $\varepsilon=1$.
Our results are robust with respect to the choice of recolouring rate $\tau_R^{-1}$ and initial conditions.

\subsection{Full Nucleus}
To model the whole nucleus we perform simulations of a melt of annealed polymers at different monomer densities $\rho=N/V$ and $\varepsilon/k_BT_L$. We here consider $N=50$ polymers with $M=256$ beads each and the range of parameters employed are $\rho=0.1$ -- $0.8$ $\sigma^{-3}$ and $\varepsilon/k_BT_L=0.75$--$1.1$. The insets of Fig.~2 in the main text are obtained using the following parameters: $\rho=0.1 \sigma^{-3}$, $\varepsilon/k_BT_L=0.7$ (Uniform Disordered); $\rho=0.7 \sigma^{-3}$, $\varepsilon/k_BT_L=0.7$ (Uniform Ordered);  $\rho=0.1 \sigma^{-3}$, $\varepsilon/k_BT_L=1.1$ (Demixed Ordered).

\section{Captions of Supplementary Movies}
\begin{itemize}
   
	\item Movie M1: Time evolution of the system described by 
eqs. \eqref{eq:modelc}, after quench $Q_1$ ($\Gamma_m = \Gamma_n = D_m = D_n = 1$, $\chi = 1$, and $n_0=2$). The system is initialised in a UD phase (homogeneous density, incoherent magnetisation), and evolves towards a UO phase, where the system is still homogeneous, but the magnetisation is organised in big clusters of coherent magnetisation.
    
	\item Movie M2: Time evolution of the system described by eqs. \eqref{eq:modelc}, after quench $Q_2$ ($\Gamma_m = \Gamma_n = D_m = D_n = 1$, $\chi = 3.5$, and $n_0=0.5$). 
Here the system is initialised in a UD phase (homogeneous density, incoherent magnetisation), and evolves towards a PDO phase, where the system organizes in clusters, and it is characterised by weak density variations.

	\item Movie M3: Time evolution of the system described by eqs. \eqref{eq:modelc}, after quench $Q_3$ ($\Gamma_m = \Gamma_n = D_m = D_n = 1$, $\chi = 6$, and $n_0=2$). The system is initialised in a UD phase (homogeneous density, incoherent colouring), and evolves towards a DO phase, where the system organizes in clusters and it is characterised by strong density variations.

	\item Movie M4: Time evolution of the system described by eqs. \eqref{eq:modelc}, following the quench $Q_4$ ($\Gamma_m = \Gamma_n = D_m = D_n = 1$, $\chi = 6$, and $n_0=0.2$). The system is initialised in a UD phase (homogeneous density, incoherent magnetisation), and evolves towards a DO phase, where the system organizes in clusters and it is characterised by strong density variations. Compared to the Movie M3, the clusters appear to be smaller.

	\item Movie M5: BD simulations of a melt of magnetic annealed polymers with monomer density $\rho=0.5 \sigma^{-3}$ and $\varepsilon/k_BT_L=0.7$.  The systems is initialised with a random colouring and it evolves towards a uniform ordered state where the large majority of beads are red via spontaneous symmetry breaking.

	\item Movie MS1: Numerical integration of eqs. \eqref{eq:modelC+switch}, with parameters  $\Gamma_m = \Gamma_n = D_m = D_n = 1$, $\chi = 4$, $n_0=2$, $\sigma_a =0.1$, and $\sigma_i = 0.5$. The system is initialised in a UD phase (homogeneous density, incoherent magnetisation). 	We observe that both the active and inactive densities organize in patterns similar to the ones observed in the PDO phase at the equilibrium. Remarkably, while the magnetisation in the equilibrium PDO phases was negligible in the low density regions, here  it assumes a positive (or negative) value that is consistent with  the neighbouring high-density areas.

	\item Movie MS2:  Numerical integration of eqs. \eqref{eq:modelC+switch}, with parameters  $\Gamma_m = \Gamma_n = D_m = D_n = 1$, $\chi = 6$, $n_0=2$, $\sigma_a =10$, and $\sigma_i = 2$. The system is initialised in a UD phase (homogeneous density, incoherent magnetisation). Both the density fields, and the magnetisation field show a behaviour similar to the one observed in the equilibrium DO phase.
	
	\item Movie MS3: Numerical integration of eqs. \eqref{eq:modelC+switch}, with parameters  $\Gamma_m = \Gamma_n = D_m = D_n = 1$, $\chi = 6$, $n_0=2$, $\sigma_a =10$, and $\sigma_i = 2$. The system is initialised in a UD phase (homogeneous density, incoherent magnetisation). Active regions ($n_a$) organizes in clusters, with strong density variations (DO phase). Note that inactive regions ($n_i$),  still forms clusters, but with low density variations (PDO phase). 
	These clusters present coherent magnetisation.
	
	\item Movie MS4: Numerical integration of eqs. \eqref{eq:modelC+switch}, with parameters  $\Gamma_m = \Gamma_n = D_m = D_n = 1$, $\chi = 6$, $n_0=2$, $\sigma_a =20$, and $\sigma_i = 50$. The system is initialised in a UD phase (homogeneous density, incoherent magnetisation), and evolves toward a UO phase (homogeneous density, coherent magnetisation).

	\item Movie MS5: BD simulations of a melt of magnetic annealed polymers with monomer density $\rho=0.8 \sigma^{-3}$ and $\varepsilon/k_BT_L=0.9$ and switching rate $\kappa = 10^{-4} \tau_B$. This Movie shows that the evolution towards a uniformly coloured state is arrested and epigenomic (epigenetic and density) domains appear. For simplicity we only show the beads that are either red or blue ($q=-1,1$) and not the neutral or inactive types. 
	
	\item Movie MS6: BD simulations of a melt of magnetic annealed polymers with monomer density $\rho=0.8 \sigma^{-3}$ and $\varepsilon/k_BT_L=0.9$ and switching rate $\kappa = 10^{-5} \tau_B$. Compared with Movie MS5, the domains appear larger. 
	For simplicity we only show the beads that are either red or blue ($q=-1,1$) and not the neutral or inactive types.

\end{itemize}

\end{document}